# Dzyaloshinskii-Moriya-type interaction and Lifshitz invariant in Rashba 2D electron gas systems

ALEXANDER P. PYATAKOV[1,2(a,b)], ANATOLY K. ZVEZDIN[2-4]

[1] *Physics Department, M.V. Lomonosov Moscow State University - Leninskie gori, MSU, 119991, Moscow, Russia*
[2] *A.M. Prokhorov General Physics Insitute of the Russian Academy of Science - Vavilova 38, Moscow, 119991, Russia*
[3] *P.N. Lebedev Physical Institute of the Russian Academy of Science - Leninskiy prospekt, 53, Moscow, 119991,Russia*
[4] *Moscow Institute of Physics and Technology State University - Institutskii per. 9, Dolgoprudny, 141700, Russia*

**Abstract** – The origin of chiral magnetic structures in ultrathin films of magnetic metals is analyzed. It is shown that the Lifshitz-type invariant term in the macroscopic thermodynamic potential can be derived from spin-orbit Rashba Hamiltonian in two dimensional electron gas (2DEG). The former is the prerequisite for existence of spin cycloid, skyrmions and other chiral phenomena observed in thin films. The estimation of the period of spin cycloid gives the value of an order of 10 nm that agrees well with the results of scanning probe microscopy observation reported in the literature.

**1.Introduction.** – Spatially modulated spin structures have been inspiring researchers for several decades due to the rich diversity of extra-functionalities they render: ferroelectricity [1–3] sensitivity to the mechanical stress [4], magnetic and electric fields [5],[6] spin currents [7],[8] etc. There are two basic mechanisms that lead to the spatial spin modulation: it can be the competing exchange interactions [9], the inhomogeneous Dzyaloshinskii interaction associated with linear in spatial derivatives contributions to the thermodynamic potential [10–12].

Of particular interest are the chiral magnetic structures that appear in thin films of magnetic metals [13], [14]. Due to the violation of the central symmetry of a bulk crystal the electron and atoms at the surface participate in antisymmetrical exchange or Dzyaloshinskii-Moriya (DM) interaction proportional to the vector product of the localized spins [$S_1 \times S_2$] that gives rise to non-uniform magnetic structures [13–16]. It should be noted, however, that the existence of the DM interaction does not always mean the formation of the spatial modulated structures. The immediate result of the DM-type exchange coupling is the microscopic canting of the spins of neighboring atoms that might have two macroscopic consequences: 1) weak ferromagnetism, i.e. the homogeneous state with noncollinear and not fully compensated magnetizations of antiferromagnetic sublattices 2) the inhomogeneous spin cycloid ordering, i.e. spatially modulated magnetic structure with modulation direction lying in the magnetization rotation plane. In the later case the existence of the DM interaction is not a sufficient but only a necessary condition for spin cycloid formation. The prerequisite of spin cycloid is the presence in the thermodynamic potential the Lifshitz–type invariant term [17] linear in spatial derivatives of the magnetic order parameter **m**: $[(\mathbf{P} \cdot \mathbf{m})div\mathbf{m} - \mathbf{P} \cdot (\mathbf{m} \cdot \nabla)\mathbf{m}]$, where **P** is polar vector. The polar direction can be pointed out by the displacement of ions in the crystals [18–20] or by the surface normal **Z** in the case of a thin magnetic film [17]. The experimental evidence of the formation of chiral structures in thin magnetic films was the observation of incommensurate spin structures: the spin cycloid with a period about 10 nm in single atomic layer of Mn [13]; the stripe domain structure with Neel type domain walls in the double atomic Fe layers of iron [14],[21],[8]; and the skyrmions [16]. These phenomena are just the aspects of the general concept of spin flexoelectricity [18] and can be understood from the analogy between the symmetry of thin films, the crystal subjected to flexural strain, the fan-shaped molecular structures in liquid crystals [22], [23], and spin cycloids in magnets [24]. However analysis of microscopic origin of spin flexoelectricity in the case of thin films is still lacking.

In this paper the mechanism that induces the chiral spin structures in 2D electronic gas (2DEG) is analyzed. It is shown that starting from a basic formula for the exchange interaction in 2DEG with Rashba spin-orbit coupling the Dzyaloshinskii-Moriya interaction described by Lifshitz invariant can be obtained.

---

(a) E-mail: `pyatakov@physics.msu.ru`; zvezdin@gmail.com
(b) Present address: Physics Department, M.V. Lomonosov MSU – Leninskie gory, 119991, Moscow, Russia.

## 2.1. RKKY interaction in 2DEG system.

Conventional approach to the microscopic analysis of origin of the chiral magnetic structures is based on the super exchange interaction [25]. The metallic systems were previously considered in the context of the three-site indirect exchange interaction via high anisotropic ligand in amorphous magnetics [26]. The presence of spin cycloid structure in thin magnetic films can be understood from the perspective of the new variety of Ruderman-Kittel-Kasuya-Yosida (RKKY) exchange coupling between localized spins [27]. The spin precession of conduction electrons in 2DEG due to the spin-orbit coupling gives rise to the so-called *twisted RKKY interaction* in a "twisted" spin space where the spin quantization axis rotates from point to point.

Starting from the conduction electron Hamiltonian with Rashba-type spin orbit coupling:

$$H_0 = -\frac{\hbar^2}{2m}\nabla^2 + \alpha(-i\hbar\nabla \times \mathbf{Z})\cdot\boldsymbol{\sigma}, \quad (1)$$

where α is the strength of the spin orbit coupling and $\boldsymbol{\sigma}$ is the vector of Pauli spin matrices, the twisted RKKY interaction for a pair of localized spins in 2DEG can be obtained [27] that takes the form similar to the conventional RKKY interaction:

$$H_{12}^{RKKY} = F(R)\mathbf{S}_1\cdot\mathbf{S}_2(\theta) \quad (2).$$

where F(R) is the range function showing the dependence of the exchange coupling on the distance, $\mathbf{S}_1$, $\mathbf{S}_2$ are the localized spins interacting via conduction electrons, R is the distance between interaction ions, $\theta$ is the angle of the rotation of spin quantization axis of the second localized spin $\mathbf{S}_2$ with respect to the one of $\mathbf{S}_1$.

Although the appearance Hamiltonian of the twisted RKKY coupling (2) and the conventional one is very similar they lead to the strikingly different consequences. Usually RKKY interaction results in the collinear (ferromagnetic or antiferromagnetic) orientation of the neighboring spins. Meanwhile the twisted RKKY interaction, that takes into account the spin precession of the conduction electrons caused by Rashba coupling, yields a noncolinear spin ordering. From the mathematical standpoint the product of spins implies not only the scalar product of the spins but also the vector product as well:

$$\mathbf{S}_1\cdot\mathbf{S}_2(\theta_{12}) = \cos\theta_{12}(\mathbf{S}_1\mathbf{S}_2) + \sin\theta_{12}[\mathbf{S}_1\times\mathbf{S}_2]_y + (1-\cos\theta_{12})S_1^y S_2^y, \quad (3)$$

where $\theta = 2k_R(x_1 - x_2)$ (spins are supposed to be located in the positions $x_1$ and $x_2$ on the *x*-axis [27]), the $k_R$ is Rashba splitting proportional to the spin orbit coupling α:

$$k_R = \frac{m\alpha}{\hbar^2}. \quad (4)$$

In the case of the ultrathin film the 2D array of atoms should be considered. For this case the cross spin product in the second term of (3) takes the form of the mixed product $([\mathbf{Z}\times\mathbf{r}]\cdot[\mathbf{S}_1\times\mathbf{S}_2])$ where **r** is the unit vector connecting the exchange coupled ions and **Z** is the unit vector normal to the plane (along the effective electric field of the Rashba spin orbit interaction).

The range function of the twisted RKKY interaction can be expressed in the following way [27]:

$$F(R) = -\frac{J^2}{2\pi^2}\frac{m}{\hbar^2}\frac{\sin(2k_F R)e^{-R/L}}{R^2}, \quad (5a)$$

$$k_F = \sqrt{\frac{2m\varepsilon_F}{\hbar^2} + k_R^2}, \quad (5b)$$

where *J* is the strength of s-d interaction, *L* is the mean free path of the conduction electrons that shows the self-damping of the exchange interaction [28],[29].

The range function dependence on the distance for Mn lattice is shown in fig. 1. It can be seen that for the neighboring atoms its value is positive. For symmetric exchange (the first term in (3)) this means anti-parallel orientation of the spins of the neighboring atoms (fig.2). However the second DM-type term in the spin product (3) results in the slight canting of their spins $\delta S \ll S$ since DM term is the first order of smallness in $\theta$ ($\theta \ll 1$ for neighboring atoms): $[\mathbf{S}_1\times\mathbf{S}_2] = [\mathbf{S}_1\times(-\mathbf{S}_1+\delta\mathbf{S})] = [\mathbf{S}_1\times\delta\mathbf{S}]$

The third term in (3) corresponds to the anisotropy of the exchange interaction. It is the second order of smallness in $\theta$ and in the following discussion it will be neglected.

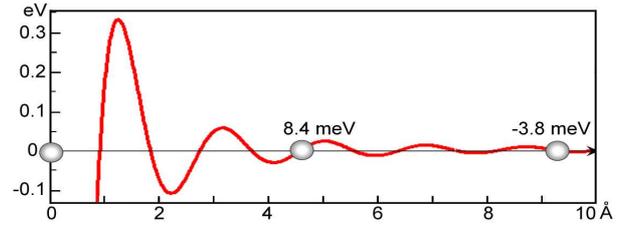

Fig. 1: (Colour on-line) The range function *F(R)* for Mn single layer: J=9.4 eV Angstroem$^2$ [13], the positions of atoms in lattice are shown with grey balls on the graph (period of the lattice is taken *a*=4.7 Angstroem). The corresponding energies are written at the atom position.

To calculate the DM-like contribution to the exchange energy per atom with a spin $\mathbf{S}_0$ we should make the summation over its neighbors. For simplicity we consider the tetragonal 2D lattice (generalization on the lower symmetry case is straightforward):

$$V_{FlexoME} = \sum_{n,m}\left(F(R_{n,m})\sin(\theta_{n,m})\cdot([\mathbf{Z}\times\mathbf{r}]\cdot[\mathbf{S}_0\times\delta\mathbf{S}_{n,m}])\right)$$

$\theta_{n,m}$ is the angle between spin quantization axes of the (0,0)-ion with spin $\mathbf{S}_0$ and the (n,m)-ion (see fig.2, inset).

The vector $\delta\mathbf{S}_{n,m}$ can be expanded in Taylor series: $\delta\mathbf{S}_{n,m} = a(\nabla_x\mathbf{S}\cdot n + \nabla_y\mathbf{S}\cdot m) + ...$, where *a* is the lattice parameter. We restrict ourselves to the first term of the expansion. The transition from the microscopic representation to the continuous one could be made:

$$V_{FlexoME} = a\sum_{n,m}\left(F(a\cdot\sqrt{n^2+m^2})\sin(\theta_{n,m})\cdot([\mathbf{Z}\times\mathbf{r}]\cdot[\mathbf{S}_0\times(\mathbf{n}\cdot\nabla)\mathbf{S}])\right) =$$
$$= V_f[(\mathbf{Z}\cdot\mathbf{S})div\mathbf{S} - \mathbf{Z}\cdot(\mathbf{S}\cdot\nabla)\mathbf{S}], \quad (6)$$

where vector **n**=(n,m), and the energy of spin flexoelectric interaction per atom:

$$V_f = a\sum_{n,m}n\cdot F(a\sqrt{n^2+m^2})\cdot\sin(\theta_{n,m})\cdot \quad (7)$$

Thus the DM-type term in the spin product (3) leads us to the Lifshitz invariant in the thermodynamical potential and, as a consequence, to the spin cycloid structure.

## 2.2. Spin cycloid in 2DEG system.

Let us assume without loss of generality that the spin cycloid is running along x-axis (fig.2). Although the spins of neighboring atoms are nearly anti-parallel they form the long range periodic structure that can be described by the slow rotation of antiferromagnetic vector $\mathbf{L}=\mathbf{S}_1-\mathbf{S}_2$.

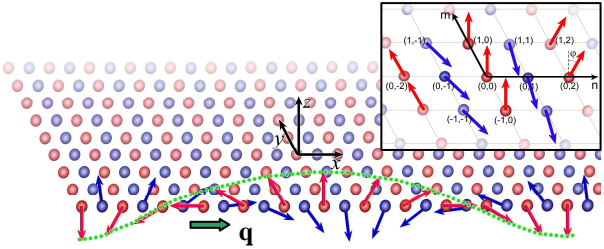

Fig. 2 (Color on-line): The cycloid structure in the single layer of magnetic metal atoms. The indexes (n,m) of neighboring atoms are shown in the inset.

The spin cycloid is described by the dependence of the angle $\varphi = \varphi(qx)$ between the antiferromagnetic vector $\mathbf{L}$ and z-axis that is normal to the film, where $\mathbf{q}$ is a wave vector of the cycloid ($\mathbf{q} \parallel \mathbf{OX}$). It leads to the spatial modulation of the relevant terms in the spin product (3):

$$\mathbf{S}_1 \cdot \mathbf{S}_n(\theta_{n,m}) = (-1)^n \cos\theta_{n,m} \cos(\varphi) + \sin\theta_{n,m} \sin(\varphi).$$

Thus the symmetric exchange can be found as:

$$H_1 = \sum_{n,m} F(a|\mathbf{n}|) \cdot \cos(\theta_{n,m}) \cdot (\mathbf{S}_1 \cdot \mathbf{S}_2) =$$
$$= \sum_{n,m} (-1)^n F(a\sqrt{n^2+m^2}) \cdot \cos(\theta_{n,m}) \cdot \cos(qan)$$

Approximating $\cos(q\,x) \approx 1 - \frac{(q\,x)^2}{2} = 1 - \frac{(na)^2(\nabla\varphi)^2}{2}$

we can obtain the surface energy density corresponding to the inhomogeneous exchange:

$$W_1 = \frac{1}{a^2}\sum_{n,m}(-1)^n F(a\sqrt{n^2+m^2}) \cdot \cos(\theta_{n,m}) \cdot \frac{(qan)^2}{2} = A(\nabla\varphi)^2 \quad (8)$$

where $A$ is the exchange stiffness

$$A = \frac{1}{2}\sum_{n,m}(-1)^{n+1} n^2 F(a\sqrt{n^2+m^2}) \cdot \cos(\theta_{n,m}). \quad (9)$$

The surface energy corresponding to the DM-like interaction can be found in analogous way for continuous approach:

$$W_2 = \gamma \cdot (\nabla\varphi) \quad (10)$$

where

$$\gamma = \frac{1}{a}\sum_{n,m} n \cdot F(a\sqrt{n^2+m^2}) \cdot \sin(\theta_{n,m}) \quad (11)$$

is 2D analogue of spin flexoelectric constant.

Let us estimate the period of the cycloid in harmonic approximation. It can be found by minimization of the energy:

$$W = A(\nabla\varphi)^2 + \gamma \cdot (\nabla\varphi) = Aq^2 + \gamma q, \quad (12)$$

$$q_0 = \frac{\gamma}{2A} = \frac{1}{a}\frac{\sum_{n,m} n \cdot F_2(a\sqrt{n^2+m^2}) \cdot \sin(\theta_{n,m})}{\sum_{n,m}(-1)^{n+1} n^2 F_2(a\sqrt{n^2+m^2}) \cdot \cos(\theta_{n,m})} \quad (13),$$

Using the values for lattice constant of Mn single layer $a$=4.7 *Angstroem* [13], and mean free path length $L$=1nm [29] the dependence of the period of the cycloid $\lambda=2\pi/q_0$ on the Rashba splitting $k_R$ (that is proportional to the spin orbit interaction (4)) can be calculated. It is quite close to the hyperbolic curves that can be explained by the fact that the exchange stiffness $A$ is nearly independent on spin orbit coupling while the constant $\gamma$ is proportional to it.

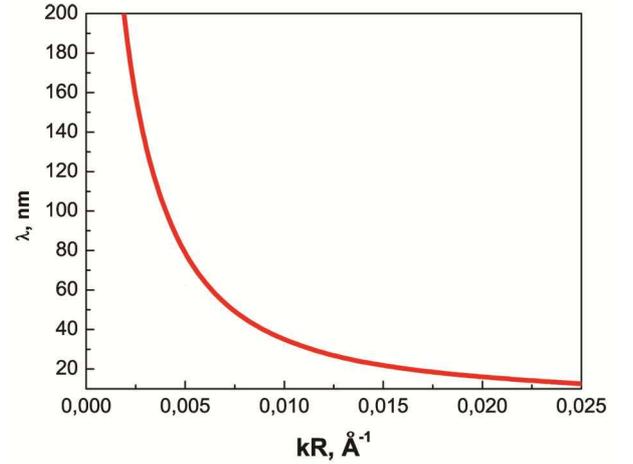

Fig. 3: (Colour on-line): The dependence of the cycloid period on $k_R$.

For values $k_R$=0.02 *Angstroem*$^{-1}$ close to the one used in [27] we obtain the spin cycloid period ~ 12nm that agrees well with experimentally founded values in [13].

## 3. Conclusion

Thus in contrast to the conventionally proposed mechanism that is based on the super-exchange model of the Dzyaloshinskii-Moriya interaction [16] in thin films another scenario can be realized originating from the twisted RKKY interaction in 2DEG system with Rashba spin-orbit coupling. This type of the antisymmetrical exchange between localized spins mediated by the conduction electrons looks more natural for the case of metallic ultrathin films than super-exchange typical for dielectrics. Note that idea of polar direction is inherent for twisted RKKY interaction. Moreover, no additional assumption (like lattice distortion or non-centrosymmetrical crystal structure) is made in this case. Thus the presence of spin cycloid as well as other chiral structures like Neel type domain wall and skyrmions is a direct consequence of the twisted RKKY interaction. It can provide new intriguing possibilities for the emergent field of flexomagnetism [30] that relates the symmetry of magnetic films and magnetochirality.

A.P. Pyatakov *et al.*

***

The work is supported by RFBR grants #13-02-12443-ofi-m, #14-02-91374 ST_a.

[1] Cheong S-W and Mostovoy M 2007 Multiferroics: a magnetic twist for ferroelectricity. *Nature materials* **6** 13–20

[2] Eerenstein W, Mathur N D and Scott J F 2006 Multiferroic and magnetoelectric materials. *Nature* **442** 759–65

[3] Khomskii D 2009 Classifying multiferroics: Mechanisms and effects *Physics* **2**

[4] Sando D, Agbelele A, Rahmedov D, Liu J, Rovillain P, Toulouse C, Infante I C, Pyatakov a P, Fusil S, Jacquet E, Carrétéro C, Deranlot C, Lisenkov S, Wang D, Le Breton J-M, Cazayous M, Sacuto A, Juraszek J, Zvezdin a K, Bellaiche L, Dkhil B, Barthélémy A and Bibes M 2013 Crafting the magnonic and spintronic response of BiFeO3 films by epitaxial strain. *Nature materials* **12** 641–6

[5] Pyatakov A P, Sechin D A, Sergeev A S, Nikolaev A V., Nikolaeva E P, Logginov A S and Zvezdin A K 2011 Magnetically switched electric polarity of domain walls in iron garnet films *EPL (Europhysics Letters)* **93** 17001

[6] Pyatakov A P, Meshkov G A and Zvezdin A K 2012 Electric polarization of magnetic textures: New horizons of micromagnetism *Journal of Magnetism and Magnetic Materials* **324** 3551–4

[7] Emori S, Bauer U, Ahn S, Martinez E and Beach G S D 2013 Current-driven dynamics of chiral ferromagnetic domain walls *Nature materials* **12** 611–6

[8] Chen G, Ma T, N'Diaye A T, Kwon H, Won C, Wu Y and Schmid A K 2013 Tailoring the chirality of magnetic domain walls by interface engineering. *Nature communications* **4** 2671

[9] Izyumov Y A 1984 Modulated, or long-periodic, magnetic structures of crystals *Soviet Physics Uspekhi* **27** 845

[10] Dzyaloshinskii I E 1964 Theory of helicoidal structures in antiferromagnets 1. Nonmetals *Journal of Experimental and Theoretical Physics* **19** 960–71

[11] Bak P and Jensen M H 1980 Theory of helical magnetic structures and phase transitions in MnSi and FeGe *Journal of Physics C: Solid State Physics* **13** L881

[12] Bogdanov A, Rößler U, Wolf M and Müller K-H 2002 Magnetic structures and reorientation transitions in noncentrosymmetric uniaxial antiferromagnets *Physical Review B* **66** 214410

[13] Bode M, Heide M, Von Bergmann K, Ferriani P, Heinze S, Bihlmayer G, Kubetzka a, Pietzsch O, Blügel S and Wiesendanger R 2007 Chiral magnetic order at surfaces driven by inversion asymmetry. *Nature* **447** 190–3

[14] Heide M, Bihlmayer G and Blügel S 2008 Dzyaloshinskii-Moriya interaction accounting for the orientation of magnetic domains in ultrathin films: Fe/W(110) *Physical Review B* **78** 140403

[15] Fert A 1990 Magnetic and Transport Properties of Metallic Multilayers *Material Science Forum* **59-60** 439–80

[16] Fert A, Cros V and Sampaio J 2013 Skyrmions on the track. *Nature nanotechnology* **8** 152–6

[17] Zvezdin A K 2002 Lifshitz surface invariant and space-modulated structures in thin films *Bull. Lebedev, Phys. Inst.* 7–12

[18] Zvezdin A K and Pyatakov A P 2012 On the problem of coexistence of the weak ferromagnetism and the spin flexoelectricity in multiferroic bismuth ferrite *EPL (Europhysics Letters)* **99** 57003

[19] Sergienko I a. and Dagotto E 2006 Role of the Dzyaloshinskii-Moriya interaction in multiferroic perovskites *Physical Review B* **73** 1–5

[20] Chizhikov V A and Dmitrienko V E 2013 Multishell contribution to the Dzyaloshinskii-Moriya spiraling in MnSi-type crystals *Physical Review B* **88** 214402

[21] Thiaville A, Rohart S, Jué É, Cros V and Fert A 2012 Dynamics of Dzyaloshinskii domain walls in ultrathin magnetic films *EPL (Europhysics Letters)* **100** 57002

[22] De Gennes P G and Prost J 1995 *The Physics of Liquid Crystals* (Oxford University Press)

[23] Beresnev L A, Blinov L M, Osipov M A and Pikin S A 1988 Ferroelectric Liquid Crystals *Molecular Crystals and Liquid Crystals Incorporating Nonlinear Optics* **158** 1–150

[24] Pyatakov A P and Zvezdin A K 2009 Flexomagnetoelectric interaction in multiferroics *The European Physical Journal B* **71** 419–27

[25] Moriya T 1960 New Mechanism of Anisotropic Superexchange Interaction *Phys. Rev. Lett.* **4** 228–30

[26] Fert A and Levy P M 1980 Role of Anisotropic Exchange Interactions in Determining the Properties of Spin-Glasses *Phys. Rev. Lett.* **44** 1538–41

[27] Bruno P and Utsumi Y 2004 Twisted exchange interaction between localized spins embedded in a one- or two-dimensional electron gas with Rashba spin-orbit coupling *Physical Review B* **69** 121303(R)

[28] De Gennes P G 1962 Interactions indirectes entre couches 4f dans les métaux de terres rares *J. Phys. Radium* **23** 510–21

[29] Takanaka K and Yamamoto B 1976 The Self-Damping of RKKY Interaction in Dilute Magnetic Alloys *physica status solidi (b)* **75** 279–88

[30] Hertel R 2013 Curvature-Induced Magnetochirality *Spin* **03** 1340009